\documentstyle[12pt,psfig]{article}

\input epsf

\newcommand{\ppp}[1]{%
        \setbox0=\hbox{#1}%
        \kern-.02em\copy0\kern-\wd0
        \kern+.04em\copy0\kern-\wd0
        \kern-.02em\raise.0217em\box0}

\newcommand{\lsim}{
 \mathrel{\setbox0=\hbox{$<$}\raise0.6ex\copy0\kern-\wd0
 \lower0.65ex\hbox{$\sim$}}}

\newcommand{\gsim}{
 \mathrel{\setbox0=\hbox{$>$}\raise0.6ex\copy0\kern-\wd0
 \lower0.65ex\hbox{$\sim$}}}

\begin{document}
%
\begin{titlepage}
\renewcommand{\thefootnote}{\fnsymbol{footnote}}
\makebox[2cm]{}\\[-1in]
\begin{flushright}
\begin{tabular}{l}
TUM/T39-97-28
\end{tabular}
\end{flushright}
\vskip0.4cm
\begin{center}
  {\Large\bf  Hard
     Exclusive Meson Production \\[0.2cm] and \\[0.3cm] Nonforward Parton 
     Distributions\footnote{Work supported in
    part by BMBF}
}\\ 

\vspace{2cm}
L.\ Mankiewicz\footnote{On leave of absence from N. Copernicus
Astronomical Center, Polish Academy of Science, ul. Bartycka 18,
PL--00-716 Warsaw (Poland)}, 
G. Piller and T. Weigl

\vspace{1.5cm}

\begin{center}
{\em Physik Department, Technische Universit\"{a}t M\"{u}nchen, \\
D-85747 Garching, Germany} 
\end{center}

\vspace{1cm}


\vspace{3cm}

\centerline{\bf Abstract}
\begin{center}
\begin{minipage}{15cm}
We present an analysis of twist-$2$, leading order QCD amplitudes for hard
exclusive leptoproduction of mesons in terms of double/nonforward 
parton distribution functions. 
After reviewing some general features of nonforward nucleon matrix
elements of twist-$2$ QCD string operators, 
we propose a phenomenological model for quark and gluon nonforward
distribution functions. 
The corresponding QCD evolution equations 
are solved in the leading logarithmic approximation for flavor
nonsinglet distributions. 
We derive explicit expressions for hard exclusive 
$\pi^0$, $\eta$, and  neutral vector meson 
production amplitudes and discuss general
features of the corresponding cross sections.
\end{minipage}
\end{center}

\end{center}
\end{titlepage}
\setcounter{footnote}{0}

\newpage

\section{Introduction}

Our current knowledge of the  sub-structure of nucleons is 
based to a large extent on high-energy scattering experiments  
which probe  its quark and gluon distribution functions.
The most prominent processes in this respect are 
deep-inelastic scattering and Drell-Yan leptoproduction.
In addition a large amount of information can be deduced from 
measurements of electromagnetic and weak form factors.  

Recently new observables, namely nonforward parton 
distribution functions, attract a great amount of interest. 
Although being discussed already some time ago 
\cite{Pioneer,Leipzig,Pioneer2},  
they were introduced in the context of the spin structure 
of nucleons recently in Ref.\cite{Ji96}. 
Nonforward parton distributions are a generalization 
of ordinary parton distributions \cite{Ji96,Rad97}.
However they are  also closely related to nucleon form factors. 
Thus they combine different aspects of the nucleon 
structure and offer new insights.

To summarize the relation of nonforward parton distributions to 
ordinary parton distributions recall that at the twist-$2$ level 
the latter  can be represented as normalized Fourier transforms of 
forward nucleon matrix elements of non-local QCD operators \cite{Collins}, 
constructed as gauge-invariant overlap of two quark or gluon fields 
separated by a light-like distance. 
Nonforward parton distributions are defined by the same 
non-local QCD operators -- just sandwiched between 
nucleon states with different momenta and eventually spin. 
Equally close  is the relation to nucleon form factors which are 
defined by nonforward matrix elements of the 
same QCD operators taken in the local limit.

Nonforward parton distributions are probed in processes 
where the nucleon target recoils elastically.
To select twist-$2$ correlations a large scale has to be involved. 
One possible process is deeply-virtual Compton scattering 
as investigated in Ref.\cite{Ji96}. 
Another promising class of reactions sensitive to 
nonforward distributions is hard exclusive leptoproduction of mesons  
(for recent works see \cite{Rad97,hood97,CFS97} 
and references therein). 
This is based on a factorization theorem which has been 
proven recently in Ref.\cite{CFS97}. 
It concerns the kinematic domain of large photon virtualities 
$Q^2 \gg \Lambda_{\rm QCD}^2$ and moderate 
momentum transfers $|t|\sim \Lambda_{\rm QCD}^2$. 
The theorem states that for incident longitudinally polarized 
photons the  meson production amplitudes can be factorized in 
a perturbatively calculable part, describing the interaction of 
the virtual photon with quarks and gluons of the target, 
and matrix elements which contain all information about  
the long-distance non-perturbative strong interaction dynamics  
in the produced meson and the nucleon target. 
The latter are nothing else than the wanted nonforward 
distribution functions. 

Data on the exclusive production of neutral vector mesons have  
been taken lately at high center of mass energies at HERA 
(for a review and references see \cite{Crit97}). 
In the measured kinematic domain the corresponding production cross 
sections are controlled by the nonforward gluon distribution of the 
target. 
The latter has been approximated in earlier investigations 
by the ordinary gluon distribution of the target nucleon, 
\cite{Rys93,FS94,FS96,MRT97}. 
The quality of this procedure is still a matter of debate 
(see e.g. Ref.\cite{FFGS97}).

At smaller center of mass energies,  
typical for current and future fixed target 
experiments at TJNAF \cite{TJNAF}, HERMES \cite{HERMES} 
and COMPASS \cite{COMPASS}, 
a similarity of the nonforward and ordinary 
gluon distribution in the  accessible kinematic 
domain is not likely. 
Furthermore also the nonforward quark distribution  
of the target will contribute in a significant way. 
Therefore information about nucleon 
nonforward matrix elements can be obtained from these experiments. 
In addition, measurements of other exclusive processes, 
like pseudoscalar or charged vector mesons production, 
will allow to access a  variety  of new nonforward 
quark and gluon distribution functions \cite{CFS97}.

In this paper we outline the calculation of hard 
exclusive leptoproduction amplitudes  for a variety 
of mesons. 
To obtain first insights in the magnitude and 
behavior of the corresponding cross sections a modeling of  
nonforward parton distributions is necessary. 
We use a simple ansatz which satisfies 
constraints due to forward 
parton distributions, form factors and discrete symmetries. 
Furthermore, we discuss the $Q^2$-dependence of 
nonforward quark distributions in the nonsinglet case 
by solving the corresponding evolution equation. 
We  are then in the position to provide baseline estimates 
for the differential cross sections for $\pi^0$, $\eta$ 
and neutral vector meson production. 

The structure of this paper is as follows: in Sec.2 we review  
the definition and general features of double and nonforward 
distribution functions. 
A simple model for double distributions is presented in 
Sec.3. In Sec.4 the $Q^2$-evolution of nonsinglet 
quark distributions is outlined. 
Exclusive meson production amplitudes are derived in Sec.5. 
After presenting results in Sec.6, we summarize in Sec.7.

\section{Double  and nonforward distribution functions}

We first recall definitions and basic features of 
double and nonforward distribution functions. 
For this purpose we stay close to the notation of 
Ref.\cite{Rad97}.
In leading twist quark and gluon distribution functions
of nucleons  are defined by forward matrix elements 
of non-local light-cone operators sandwiched between 
nucleon states \cite{Collins}: 
\begin{eqnarray}      \label{eq:quark_distr}
\left \langle N(P,S)  \right| 
\hat O(0,z) \left| N(P,S)\right \rangle_{z^2 = 0} 
\sim 
\int_0^1 \!\!dx  
\left[
e^{-i x (P\cdot z)} f(x) 
\pm 
e^{i x (P\cdot z)} \bar f(x)
\right].
\end{eqnarray}
Here $|N(P,S) \rangle$ represents  a nucleon  with momentum 
$P$ and spin $S$. 
The operator  $\hat O(0,z)$ stands for a product of two quark or 
gluon fields, respectively, separated by a light-like distance 
$z \sim n$,  with $n\cdot a = a^+ = a^0 + a^3$ for an  
arbitrary vector $a$.  
In Eq.(\ref{eq:quark_distr}) we suppress any dependence 
on the normalization scale. 
Note that the sign in front of the anti-parton distribution $\bar f$ 
is determined by the charge conjugation property 
of the operator $\hat O$ in the local limit.
Forward matrix elements as in  Eq.(\ref{eq:quark_distr}) 
are explored in processes where no momentum is transfered 
to the target, e.g. in forward virtual Compton scattering or 
deep-inelastic scattering, respectively.

\subsection{Double distribution functions}

In exclusive production processes a non-vanishing  
momentum $r = P-P'$ is  transfered to the target. 
Thus in leading twist rather nonforward matrix elements of 
quark and gluon light-cone operators are probed. 
These can be parametrized through so called 
``double distribution functions'' which where introduced in 
Ref.\cite{Rad97}. 
For the unpolarized quark distribution one has: 
\begin{eqnarray} \label{eq:dd_quark}
&&\hspace{-0.5cm}
\left \langle N(P',S')\right| 
\bar\psi(0) \hat z \left[0,z\right] \psi(z) 
\left|N(P,S)\right \rangle_{z^2 = 0} 
\!=\!
\nonumber\\
&&\hspace{-0.5cm}
\bar N(P',S')\,\hat z \,N(P,S) 
\!\int_0^1 \!\!dx \!\int_0^{\bar x} \!\! dy\! 
\left[
e^{-i x (P\cdot z) - i y(r\cdot z)} F(x,y,t) 
- 
e^{i x (P\cdot z) - i \bar y(r\cdot z)} \bar F(x,y,t)
\right] 
\nonumber \\
&&\hspace{-0.5cm} + 
\bar N(P',S')\frac{\sigma_{\mu\nu}z^{\mu}r^{\nu}}{i M} N(P,S) 
\!\int_0^1 \!\!dx \!\int_0^{\bar x} \!\! dy\! 
\left[
e^{-i x (P\cdot z) - i y(r\cdot z)} K(x,y,t) 
- 
e^{i x (P\cdot z) - i \bar y(r\cdot z)} \bar K(x,y,t)
\right], 
\nonumber \\
\end{eqnarray}
where we use the notation $\hat z = z_{\mu} \gamma^{\mu}$. 
The operator on the LHS of Eq.(\ref{eq:dd_quark}) is built from quark fields 
connected by  the path-ordered exponential  
$[0,z] = {\cal P} \mbox{exp} [ -i g z_{\mu} 
\int_0^1 d\lambda A^{\mu}(z \lambda)]$ 
which guarantees gauge invariance and reduces to 
one in axial gauge $n\cdot A=0$ 
($g$ stands for the strong coupling constant and 
$A^{\mu}$ denotes the gluon field). 
On the RHS we have the product of the  Dirac spinors  
for the initial and final nucleon.
The quark and antiquark double distribution functions 
$F$, $\bar F$, $K$ and $\bar K$  depend on the momentum transfer $t = r^2$,
and two light-cone variables $x$ and $y$  
($\bar x = 1-x,\,\bar y = 1-y$).
Their interpretation is discussed at length in 
Ref.\cite{Rad97}. 
Note that 
the distributions $K$ and $\bar K$ 
enter   proportional to the momentum transfer $r$. 

A comparison of Eqs.(\ref{eq:quark_distr}) and (\ref{eq:dd_quark})
demonstrates that in the forward limit, $r \rightarrow 0$, 
the double distributions reduce to ordinary 
quark distributions, e.g.: 
\begin{equation}
f(x) = \int_0^{\bar x} \!\! dy \,F(x,y,t=0).
\label{eq:nonf_to_forw}
\end{equation}
For the polarized quark and the unpolarized  gluon distribution 
one has \cite{Rad97}: 
\begin{eqnarray} 
&&
\left \langle N(P',S')\right| 
\bar\psi(0) \hat z \gamma^5\left[0,z\right] \psi(z) 
\left|N(P,S)\right \rangle_{z^2 = 0} 
\!=\!
\nonumber\\
&&
\bar N(P',S')\hat z \gamma^5 N(P,S) 
\!\int_0^1 \!\!dx \!\int_0^{\bar x} \!\! dy\! 
\left[
e^{-i x (P\cdot z) - i y(r\cdot z)} \Delta F(x,y,t) 
+ 
e^{i x (P\cdot z) - i \bar y(r\cdot z)} \Delta \bar F(x,y,t)
\right] \nonumber \\
&& + K-\mbox{terms},\nonumber \\
\label{eq:dd_quarkpol} \\
&&
z_\mu z_\nu
\left \langle N(P',S')\right| 
{\mbox {Tr}}\, G^{\mu\xi}(0)[\Delta;0]
{G}_\xi\,^\nu(z),
\left|N(P,S)\right \rangle_{z^2 = 0}
\!=\!
\nonumber\\
&&
\bar N(P',S')\,\hat z \,N(P,S) \frac{\bar P \cdot z}{4}  
\!\int_0^1 \!\!dx \!\int_0^{\bar x} \!\! dy\! 
\left[
e^{-i x (P\cdot z) - i y(r\cdot z)} +  
e^{i x (P\cdot z) - i \bar y(r\cdot z)}\right]
G(x,y,t) \nonumber \\
&& + K-\mbox{terms}, 
\label{eq:dd_gluon}
\end{eqnarray}
with $\bar P = (P'+P)/2$. 
In Eqs.(\ref{eq:dd_quarkpol},\ref{eq:dd_gluon}) 
we have only indicated the so-called "$K$-terms"
which are  proportional to the momentum transfer $r$ as in  
Eq.(\ref{eq:dd_quark}). 
Of course, relations similar to (\ref{eq:nonf_to_forw})
hold also for the double distributions 
in Eqs.(\ref{eq:dd_quarkpol},\ref{eq:dd_gluon}).

It is important to realize that the double distributions defined above fulfill
a symmetry constraint based on hermiticity.
For the unpolarized quark distributions $F$ and $\bar F$ this can be seen by
considering the C-even ($\Sigma$) and C-odd ($\Delta$) combinations:
\begin{equation}
\left.
\begin{array}{l} 
\Sigma \\ 
\Delta
\end{array}
\right\}
\equiv 
\left \langle P',S'\right| 
\bar\psi\left(\frac{z}{2}\right) \hat z \left[\frac{z}{2},-\frac{z}{2}\right] 
\psi\left(-\frac{z}{2}\right) 
\left|P,S\right \rangle_{z^2 = 0}  
\pm \left(z \longleftrightarrow -z \right).
\end{equation}
{}From the definition (\ref{eq:dd_quark}), combined with the fact that 
double distributions are real, it follows  that  
$\Sigma$ and $\Delta$ are purely imaginary and real,
respectively:
\begin{eqnarray}
\Sigma &=& 2 i \bar N(P',S')\,\hat z \,N(P,S) 
\nonumber \\
&&\times
\!\int_0^1 \!\!dx \!\int_0^{\bar x} \!\! dy\! 
\left[
F(x,y,t) 
+ 
\bar F(x,y,t)
\right]
\sin\left[ x \bar P\cdot z + r \cdot z \left(y - \frac{1-x}{2}\right)
\right] 
+ K-\mbox{terms},\nonumber \\
\Delta &=& 2 \bar N(P',S')\,\hat z \,N(P,S) 
\nonumber \\
&&\times
\!\int_0^1 \!\!dx \!\int_0^{\bar x} \!\! dy\! 
\left[
F(x,y,t) 
- 
\bar F(x,y,t)
\right]
\cos\left[ x \bar P\cdot z + r \cdot z \left(y - \frac{1-x}{2}\right)
\right] 
+ K-\mbox{terms}. \nonumber \\
\end{eqnarray}
As a consequence $\Sigma = (\Sigma - \Sigma^\dagger)/2$ as well as
$\Delta=(\Delta + \Delta^\dagger)/2$ is symmetric with respect to an exchange
of the initial and final nucleon momentum, i.e. $P\leftrightarrow P'$ and
simultaneously $r \leftrightarrow - r$.  This yields the following sum rules:
\begin{eqnarray} 
&&\int_0^1 \!\!dx \!\int_0^{\bar x} \!\! dy\! 
\left[
F(x,y,t) 
+ 
\bar F(x,y,t)
\right]
\cos\left(x \bar P\cdot z\right) 
\sin\left[ r \cdot z \left(y - \frac{1-x}{2}\right)
\right] = 0, 
\\
&&\int_0^1 \!\!dx \!\int_0^{\bar x} \!\! dy\! 
\left[
F(x,y,t) 
- 
\bar F(x,y,t)
\right]
\sin\left(x \bar P\cdot z\right) 
\sin\left[ r \cdot z \left(y - \frac{1-x}{2}\right)
\right] = 0. 
\end{eqnarray}
For the polarized quark nonforward distributions as well as for 
the gluon distribution similar constraints can be derived. 
They imply that the double distributions in 
Eqs.(\ref{eq:dd_quark},\ref{eq:dd_quarkpol},\ref{eq:dd_gluon}) 
are symmetric with respect to an exchange of 
\begin{equation} \label{eq:constraint}
y \longleftrightarrow 1 - x - y.
\end{equation}
This symmetry, besides being important for  modeling  double distribution
functions, is crucial for establishing proper analytical properties of
meson production amplitudes.

In the forward limit, $r \rightarrow 0$,
Eqs.(\ref{eq:dd_quark},\ref{eq:dd_quarkpol},\ref{eq:dd_gluon})
immediately reduce to the definitions of  ordinary 
twist-two parton distributions (\ref{eq:quark_distr}).
On the other hand in the limit $z\rightarrow 0$  
Eqs.(\ref{eq:dd_quark},\ref{eq:dd_quarkpol},\ref{eq:dd_gluon}) define 
familiar nucleon form factors.  
One therefore obtains the following sum rules for double parton
distributions \cite{Pioneer2,Ji96,Rad97}: 
\begin{eqnarray} \label{eq:ff_1}
&&\int_0^1 dx \int_0^{\bar x} dy \left[
F(x,y,t) 
-
\bar F(x,y,t)
\right] = F_1(t),
\\
&&\int_0^1 dx \int_0^{\bar x} dy \left[
K(x,y,t) 
-
\bar K(x,y,t)
\right] = F_2(t),
\\
&&\int_0^1 dx \int_0^{\bar x} dy \left[
\Delta F(x,y,t) 
+
\Delta \bar F(x,y,t)
\right] = G_A(t),
\\
&&\int_0^1 dx \int_0^{\bar x} dy \left[
G(x,y,t) 
+
\bar G(x,y,t)
\right] = {\cal G}(t).
\label{eq:formfactors}
\end{eqnarray}
$F_1$, $F_2$ and $G_A$ are the Dirac, Pauli and axial-vector 
form factors of the nucleon. 
The gluon form factor ${\cal G}(t)$ is experimentally not observable. 
Its dependence on the momentum transfer has been 
estimated within QCD sum rules \cite{BGMS93}, and is determined 
by a characteristic radius of the order of $0.4 - 0.5$ fm.
Similar sum rules hold for the remaining distributions 
\cite{Ji96}.

\subsection{Nonforward distribution functions}

In the definitions of double distribution functions 
(\ref{eq:dd_quark},\ref{eq:dd_quarkpol},\ref{eq:dd_gluon})
the variable $y$ always enters in the combination 
$x + y \, \zeta \equiv X$ with $\zeta = \frac{r\cdot z}{P \cdot z}$.  
As a consequence it is possible to define e.g., unpolarized nonforward 
quark and antiquark distribution functions \cite{Rad97}:
\begin{eqnarray}
&& \hspace{-0.5cm} 
F_{\zeta}(X,t) =  \int_0^1 dx \int_0^{\bar x} dy \,F(x,y,t)\, 
\delta\left(X - (x + \zeta y)\right) \nonumber \\ 
&& =  \Theta(X\ge\zeta) \int_0^{\bar X/\bar\zeta} dy \, 
F(X-y\zeta,y,t) 
+ 
\Theta(X\le\zeta) \int_0^{X/\zeta} dy \, 
F(X-y\zeta,y,t), \nonumber \\
&&\hspace{-0.5cm} 
{\bar F}_{\zeta}(X,t) =  \int_0^1 dx \int_0^{\bar x} dy \,{\bar F}(x,y,t)\, 
\delta\left(X - (x + \zeta y)\right) \nonumber \\ 
&& =  \Theta(X\ge\zeta) \int_0^{\bar X/\bar\zeta} dy \, 
{\bar F}(X-y\zeta,y,t) 
+ 
\Theta(X\le\zeta) \int_0^{X/\zeta} dy \, 
{\bar F}(X-y\zeta,y,t),
\label{def:F_zeta}  
\end{eqnarray}
and similar for the polarized quark and gluon case.
Since the scattered nucleon is on its mass shell, i.e. $P'^2 = M^2$ 
with the nucleon mass $M$, one finds $\zeta \le 1$. 
Together with the kinematic constraint  $x + y \le 1$ 
this gives $0\le X\le 1$.  
The variable $X$ can be identified with the nucleon light-cone momentum 
fraction of the parton being removed  by the operator $\hat O$ 
from the target, while the light-cone momentum fraction of the returning 
parton is equal to $X-\zeta$. 

In terms of the nonforward distribution functions the 
light-cone correlators in 
(\ref{eq:dd_quark},\ref{eq:dd_quarkpol},\ref{eq:dd_gluon}) 
read: 
\begin{eqnarray}
&&\hspace{-0.5cm}
\left \langle N(P',S')\right| 
\bar\psi(0) \hat z \left[0,z\right] \psi(z) 
\left|N(P,S)\right \rangle_{z^2 = 0} 
\!=\!
\nonumber\\
&&
\bar N(P',S')\,\hat z \,N(P,S) 
\!\int_0^1 \!\! dX 
\left[
e^{-i X (P\cdot z)} F_{\zeta}(X,t) 
-
e^{i (X -\zeta) (P\cdot z)} \bar F_{\zeta}(X,t)
\right]
\nonumber \\
&& + 
\bar N(P',S')\frac{\sigma_{\mu\nu} z^{\mu} r^{\nu}}{iM} N(P,S) 
\!\int_0^1 \!\! dX 
\left[
e^{-i X (P\cdot z)} K_{\zeta}(X,t) 
-
e^{i (X -\zeta) (P\cdot z)} \bar K_{\zeta}(X,t)
\right], \nonumber \\ \nonumber \\
&&\hspace{-0.5cm}
\left \langle N(P',S')\right| 
\bar\psi(0) \hat z \gamma^5\left[0,z\right] \psi(z) 
\left|N(P,S)\right \rangle_{z^2 = 0} 
\!=\!
\nonumber\\
&&\hspace{0.5 cm}
\bar N(P',S')\hat z \gamma^5 N(P,S) 
\!\int_0^1 \!\! dX 
\left[
e^{-i X (P\cdot z)} \Delta F_{\zeta}(X,t) 
+ 
e^{i (X -\zeta) (P\cdot z)} \Delta \bar F_{\zeta}(X,t)
\right]
+ K-\mbox{terms},\nonumber \\
\nonumber \\
&&\hspace{-0.5cm}
z_\mu z_\nu
\left \langle N(P',S')\right| 
{\mbox{Tr}}\, G^{\mu\xi}(0)[\Delta;0]
{G}_\xi\,^\nu(z),
\left|N(P,S)\right \rangle_{z^2 = 0} 
\!=\!
\nonumber\\
&&\hspace{0.5 cm}
\bar N(P',S')\,\hat z \,N(p,s) \frac{\bar P \cdot z}{4} 
\!\int_0^1 dX 
\left[
e^{-i X (P\cdot z)} +  
e^{i (X -\zeta) (P\cdot z)}\right]
G_{\zeta}(X,t)
+ K-\mbox{terms}. 
\label{def:corr_F_zeta}
\end{eqnarray}
\\
Note that in the literature also a second parametrization of 
nonforward distributions is popular \cite{Ji96}. 
It is based on a different asymmetry parameter $\xi$:
\begin{equation}
\xi = \frac{1}{2} \frac{r \cdot z}{{\bar P} \cdot z} =
\frac{\zeta}{2-\zeta},\quad \mbox{with}\quad 0 \le \xi \le 1.
\label{def:xi}
\end{equation}
Using for example for the unpolarized quark and antiquark 
distributions
\begin{eqnarray}
F^\prime(u,\xi,t) & = & \int_0^1 dx \int_0^{\bar x} F(x,y,t) 
\delta\left(u - (x+\xi(x+2y-1))\right),  \nonumber \\
{\bar F}^\prime(u,\xi,t) & = & \int_0^1 dx \int_0^{\bar x} {\bar F}(x,y,t) 
\delta\left(u + (x+\xi(x+2y-1))\right),
\label{def:F_xi}  
\end{eqnarray}
allows to define an alternative nonforward distribution as:
\begin{eqnarray}
H(u,\xi,t) & = & \Theta(u \ge \xi) F^\prime(u,\xi,t) +
\Theta(\xi \ge u \ge -\xi) (F^\prime(u,\xi,t) -  {\bar F}^\prime(u,\xi,t)) 
\nonumber \\
&-& \Theta(-\xi \ge u) {\bar F}^\prime(u,\xi,t). 
\label{def1:F_xi}  
\end{eqnarray}
The parametrization of the nonforward matrix element
(\ref{eq:dd_quark}) in terms of the new distributions reads 
\cite{Ji96}: 
\begin{eqnarray}
&&\hspace*{-1cm} 
\left \langle N(P',S')\right| 
\bar\psi(0) \hat z \left[0,z\right] \psi(z) 
\left|N(P,S)\right \rangle_{z^2 = 0} 
= 
\nonumber \\
&&
\bar N(P',S')\,\hat z\, N(P,S) e^{-i \frac{r\cdot z}{2}}
\!\int_{-1}^1 \!\! du \,H(u,\xi,t) e^{-i u({\bar P}\cdot z)}
 + K-\mbox{terms}.
\label{def:corr_F_xi}
\end{eqnarray}
Both definitions of
nonforward parton distributions are of course equivalent.
In the following  we will always parameterize nonforward
matrix elements as in Eqs.(\ref{def:F_zeta},\ref{def:corr_F_zeta}), 
i.e. in terms of variables $X$ and $\zeta$. 

\section{Model}

Nonforward parton distributions have not yet been measured. 
Therefore one has to rely on models \cite{Models} 
in order to provide estimates for exclusive production 
cross sections. 
To guarantee a proper analytic behavior of the involved 
amplitudes,  as given by dispersion relations \cite{Col77}, 
it is favorable to model double distributions  
(\ref{eq:dd_quark},\ref{eq:dd_quarkpol},\ref{eq:dd_gluon}) 
instead of nonforward distributions.
We parametrize the former, denoted generically by $F(x,y,t;\mu_0^2)$,  
such that they  fulfill all constraints we are aware of. 
As a model ansatz at some low normalization 
scale $\mu_0^2$ we take \cite{private}:
\begin{equation} 
  F(x,y,t;\mu_0^2)  =  \frac{f(x,\mu_0^2)}{(1-x)^3} \,  h(x,y) \,  f(t) ,
\label{def:nf_part_dist}
\end{equation}
where $f(x,\mu_0^2)$ stands for the corresponding ordinary quark and 
gluon distribution, respectively. 
It is clear that one can model in this way
only the  distributions $F, \bar F, \Delta F, \Delta \bar F$ and $G$.
The "$K$-terms" in 
(\ref{eq:dd_quark},\ref{eq:dd_quarkpol},\ref{eq:dd_gluon}) 
remain unconstrained
because the corresponding forward parton distributions are not known. 
We choose:
\begin{equation}
   h(x,y)  = 6 \, y \, (1 - x - y) \label{poss},
\label{hdef}
\end{equation}
such that $F(x,y,t;\mu_0^2)$ satisfies the symmetry
constraint in Eq.(\ref{eq:constraint}).  Of course there are many possible
choices for $h(x,y)$, e.g.  $h(x,y) = 12 \left[ y - \frac{1-x}{2} \right]^2$.
The latter however results in a nonforward parton distribution which is not
continuously differentiable at $X = \zeta$.  Although we are not aware of any
principle which forbids such a behavior, we use in the following the
parametrization from Eq.(\ref{hdef}) leading to nonforward distributions 
smooth in $X$.

The form factor $f(t)$ in Eq.(\ref{def:nf_part_dist}) 
is responsible for the $t$-dependence of  the double distribution 
functions.
Motivated by the relationship between double distributions and
nucleon form factors in Eqs.(\ref{eq:ff_1} -- \ref{eq:formfactors}) 
we assume:
\begin{equation}
   f(t) = \left( \frac{1}{1 - t/\Lambda^2} \right)^2. 
\end{equation}
For quark distributions we take $\Lambda$ from fits to the nucleon vector and
axial form factors \cite{Weise}.  
It should however be mentioned 
that the meson production amplitudes, as derived in Sec.5,   
are controlled by nonforward distribution functions 
with different C-parity as compared to the above 
nucleon form factors.  
In the case of the gluon distribution the scale $\Lambda$ 
can be  taken from the QCD sum rule estimate in Ref.\cite{BGMS93}.
One should keep in mind that a factorization of the $t$-dependence 
into a form factor $f(t)$ is reasonable only at small values of $|t|$.
At large $|t|$ the  minimal component of the nucleon 
light-cone wave function dominates \cite{BroLep}. 
This should lead to a change of the entire $x$ and $y$
dependence of $F$ which cannot be absorbed 
in a pre-factor.
However, as long as we are interested in the small-$|t|$ region, the
parametrization (\ref{def:nf_part_dist}) should be acceptable from a
phenomenological point of view.

\begin{figure}[h]
\centering{\ \psfig{figure=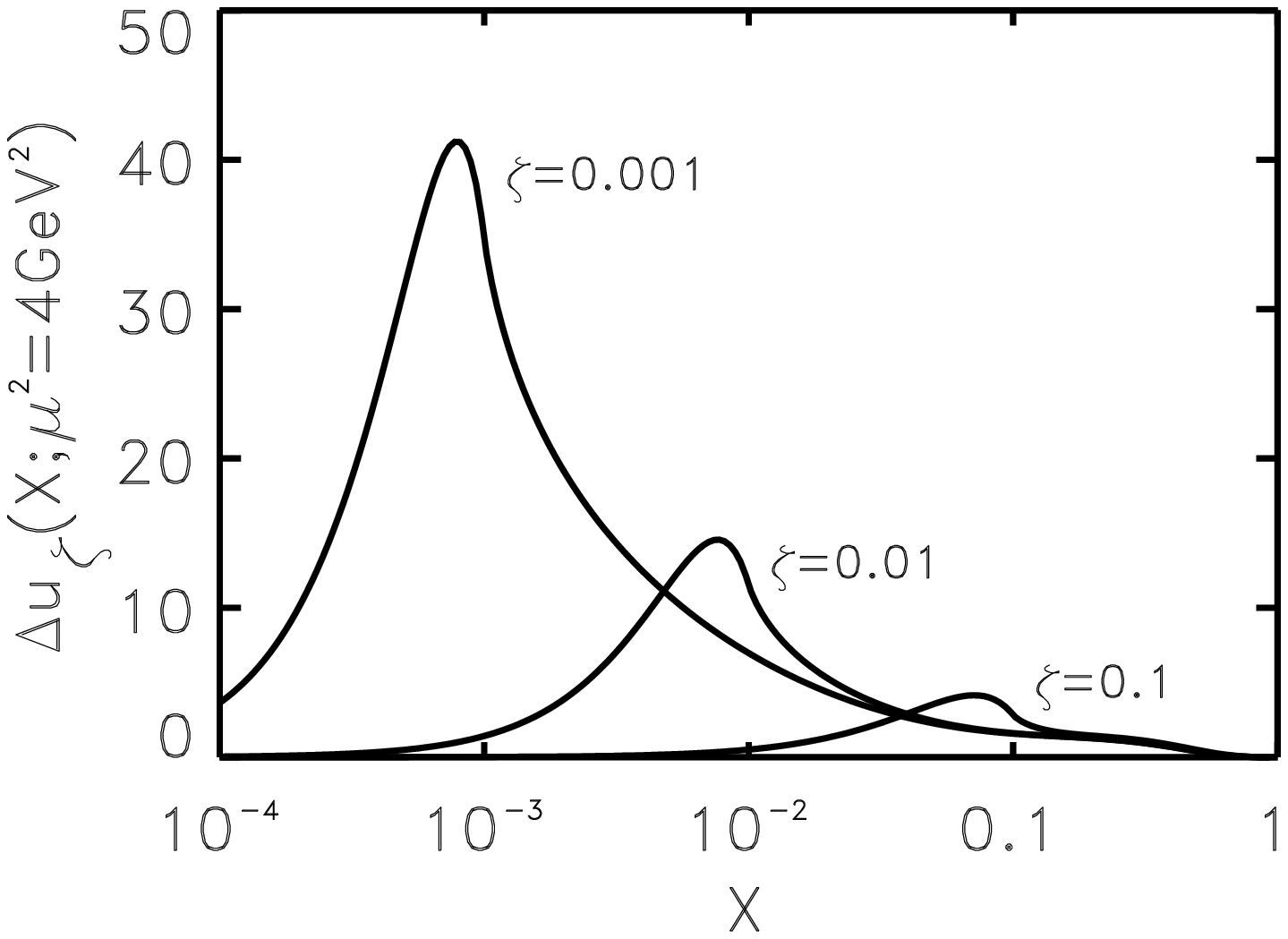,height=10.5cm}}
\caption{Polarized $u$-valence distribution
        ${\Delta u}_{\zeta}(X;\mu^2 = 4)$ from Eq.(\ref{def:nf_part_dist})  
         for different values of $\zeta$.}
\label{asyfigure}
\end{figure}

In Fig.\ref{asyfigure} we present our phenomenological model 
for the nonforward polarized 
$u$-quark distribution, ${\Delta u}_{\zeta}(X,\mu_0^2)$, taken at
$t=0$.  Here and in the following we use parametrizations for the 
corresponding forward distributions, which enter 
through Eq.(\ref{def:nf_part_dist}), from Ref.\cite{GS96}.
One can recognize the characteristic distribution amplitude like
shape for $X < \zeta$, and the similarity to ordinary 
parton distributions at $X > \zeta$.

\section{QCD evolution of nonforward distributions}

An important ingredient for  the discussion of nonforward  distribution
functions is their evolution, or dependence on the normalization scale. 
In QCD it is governed by renormalization group equations which are direct
generalizations of the well known DGLAP equations for ordinary parton
distributions.

Instead of looking at the distribution functions themselves 
it is convenient  to
consider QCD evolution at the operator level \cite{Leipzig,Evolop}. Evolution
equations for various quantities, like evolution equations
for meson distribution amplitudes \cite{BroLep,Rad}, or the DGLAP equations for
ordinary parton distributions, follow then by taking appropriate operator
matrix elements. 
At the one loop level evolution equations for operators can
be solved by an expansion in a set of multiplicatively renormalizable 
operators which are 
determined through  the conformal symmetry of one-loop
QCD \cite{Conformal}. 

In the following we outline a convenient
numerical method to perform the QCD evolution of nonforward  parton
distributions. Details of the method and a thorough discussion of its
accuracy will be given in a separate publication.
In the nonsinglet case the Gegenbauer moments of  nonforward 
parton distributions
\begin{equation}
   {\cal C}_\zeta(n;\mu_0^2) = \int_0^1 dX \,C_n^{3/2}(2\, X/\zeta - 1)
                           \, {\cal F}_{\zeta}(X;\mu_0^2)
\label{eq:gegmom}
\end{equation}
are multiplicatively renormalizable \cite{Rad97}, and their evolution reads:
\begin{equation}
    {\cal C}_\zeta(n;\mu^2) = \left[ \frac{\alpha_s(\mu^2)}{\alpha_s(\mu_0^2)}
                            \right]^{\frac{\gamma_n}{2 \beta_0}}
                            {\cal C}_\zeta(n;\mu_0^2),
    \label{gegenbauer}
\end{equation}
with $\beta_0 = 11 - \frac{2}{3} N_f$, The nonsinglet anomalous 
dimensions $\gamma_n$ are given as usual by: 
\begin{equation}
 \gamma_n = 4 \, C_f \left[ \frac{1}{2} - \frac{1}{(n+1)(n+2)}
            + 2 \sum_{j = 2}^{n+1} \frac{1}{j} \right].
\end{equation}
Because the Gegenbauer polynomials $C_n^{3/2}$ form an 
orthogonal set on the segment
$[-1,1]$, the straightforward inversion of Eq.(\ref{eq:gegmom}) is possible
only for $\zeta =1$.  On the other hand one can expand the Mellin moments of
the nonforward  parton distribution
\begin{equation}
M_n(\zeta;\mu_0^2) \equiv \int_0^1 dX \, X^n
\, {\cal F}_{\zeta}(X;\mu_0^2)
\end{equation}
in terms of  Gegenbauer moments ${\cal C}_\zeta(n;\mu_0^2)$, and evolve the
latter according to Eq.(\ref{gegenbauer}).
The corresponding expression has been  obtained in Ref.\cite{Rad97}: 
\begin{eqnarray}
   M_n(\zeta;\mu^2) & = &\zeta^n \, n! \, (n+1)! \sum_{k = 0}^n
                  \frac{(-1)^k}{2 \, \zeta^k k! \, (k+1)!} \, 
                  M_k (\zeta;\mu_0^2)
                    \nonumber \\
                    &\times&
  \sum_{l = k}^n \frac{(-1)^l \, 2 \, (2 \, l +3) (k+l+2)!}
                    {(n+l+3)! \, (n-l)! \, (l-k)!} 
                    \left[ \frac{\alpha_s(\mu^2)}{\alpha_s(\mu_0^2)}
                    \right]^{\frac{\gamma_l}{2 \beta_0}}.
\label{evolution}
\end{eqnarray}
To obtain the evolved distribution from its moments one usually performs an 
inverse Mellin transformation. In the present case, however, 
Eq.(\ref{evolution}) has a simple form only for integer values of $n$, and
therefore usual contour integration methods (see e.g. \cite{Wei96}) in the
complex $n$-plane are difficult to implement. This problem can be 
circumvented by expanding the nonforward parton distribution
$F_\zeta(X;\mu^2)$ in terms of e.g., shifted Legendre polynomials 
$P_k(2\, X - 1)$ \cite{Kry69} which are orthogonal on the  segment $[0,1]$:
\begin{equation}
F_\zeta(X;\mu^2) = \sum_{k=0}^{\infty}
                    \, (2k + 1) \,\, c_k(\zeta;\mu^2) \, P_k(2\, X - 1)  .
\end{equation}
The coefficients $c_k(\zeta;\mu^2)$ can be determined
from the set of moments $M_n(\zeta;\mu^2)$: 
if a shifted Legendre polynomial $P_k(2\, X - 1)$ follows an expansion
\begin{equation}
   P_k(2\, X - 1) = \sum_n a_k^n \, X^n,
\end{equation}
the coefficients $c_k(\zeta;\mu^2)$ are given by:
\begin{equation}
   c_k(\zeta;\mu^2) = \sum_n a_k^n \, M_n(\zeta;\mu^2) \, .
\end{equation}

\begin{figure}[h]
\centering{\ \psfig{figure=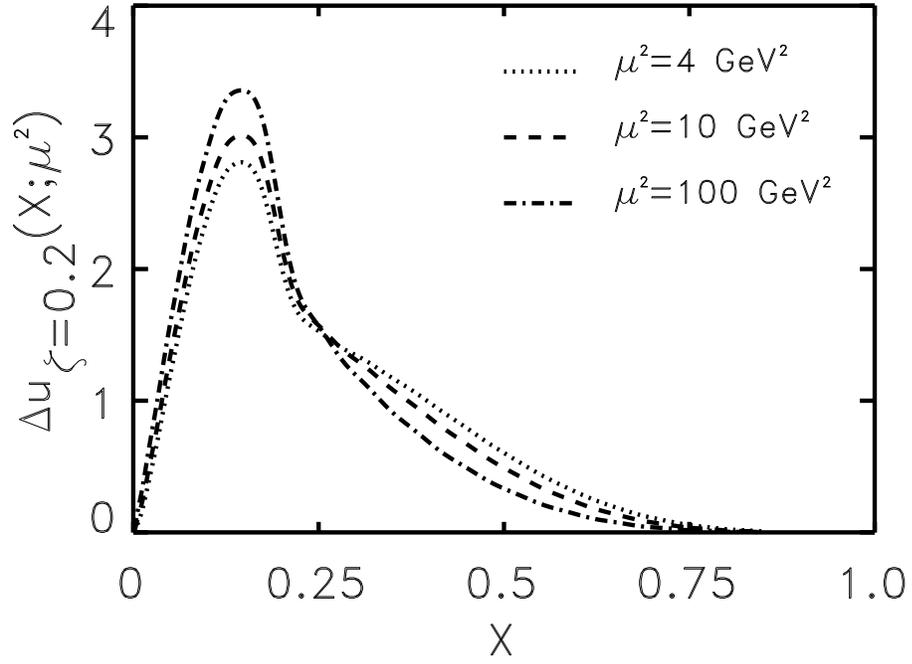,height=10.5cm}}
 \caption{Evolution of the nonforward polarized $u$-valence distribution
          ${\Delta u}_{\zeta = 0.2}(X;\mu^2)$ for different values of
          $\mu^2$.}
\label{evofigure}
\end{figure}

In Fig.\ref{evofigure} we present typical results for the evolution of the
nonforward polarized $u$-valence quark distribution, parametrized at $\mu_0^2 =
4$ GeV$^2$ according to our model in Sec.3.  For the leading order evolution we
use $N_f = 4$ and $\Lambda_{QCD} = 250$ MeV.  The evolved distribution behaves
as expected: the area under the curve remains constant as a consequence of the
vanishing anomalous dimension $\gamma_0$. The evolution shifts the distribution
to smaller values of $X$, approaching very slowly the asymptotic shape ${\Delta
  u}_{\zeta}(X; \mu^2 \rightarrow \infty) = 6 \, {\cal C}(0) \, X/\zeta^2 \, (1
- X/\zeta)$ \cite{Rad97}.  \footnote{Recently a similar method has been
  proposed independently in Ref.\cite{Reg97}.}

\section{Exclusive meson production amplitudes} 
       
A solid QCD description of hard exclusive meson production is based on
the factorization of long- and short-distance dynamics which 
has been proven in 
Ref.\cite{CFS97}. The main result is that the amplitudes for 
the production of mesons from longitudinally polarized photons 
can be split into three parts:
the perturbatively calculable hard photon-parton interaction arises from  
short
distances, while the long-distance dynamics can be factorized in terms of
nonperturbative meson distribution amplitudes and nucleon nonforward parton
distributions. In the following we derive production
amplitudes for a variety of neutral mesons in terms of these building blocks, 
staying at leading order in the strong coupling constant $\alpha_s$.

The kinematics is defined as follows:
the initial and scattered nucleon carries a  momentum  
$P$ and $P' = P- r$, respectively,  while $t = r^2$ 
stands for the squared momentum transfer. 
The momentum of the exchanged virtual photon is 
denoted as usual by $q$ with $Q^2 = - q^2$. 
Then $q' = q+r$ is the momentum of the produced meson.
Calculating the hard subprocess to leading twist accuracy 
allows to  neglect 
the momentum transfer $t$, the invariant mass of the 
nucleon target $P^2 = P'^2$, and the mass of the 
produced vector meson $q^{'2}$  
as compared with the virtuality of the photon $Q^2$.
Then the Bjorken scaling variable $x_{Bj}= Q^2/2 P\cdot q$ 
coincides with the longitudinal momentum transfer, i.e.  
$\zeta \equiv r\cdot n/P\cdot n = x_{Bj}$.

In leading order in the electromagnetic interaction we obtain 
for the S-matrix in question:  
\begin{eqnarray}
  S & = & 
  \langle M(q')\, N(P') | \hat{S} | \gamma^*_L(q) \, N(P) \rangle,  \nonumber \\
   && \nonumber \\ 
   & = & - i \int \! d^4 x \, e^{-i q\cdot x} \,\,
  \langle M(q')\, N(P') | \epsilon_L \cdot J(x) |  N(P) \rangle .
\end{eqnarray}
Here $|\gamma^*_L(q) \, N(P)\rangle$ denotes the incoming photon and 
nucleon while $\langle M(q') \, N(P')|$ represents  the final 
meson and nucleon. 
The polarization vector of the longitudinally polarized virtual photon is 
given by $\epsilon_L^\mu = \frac{i}{{Q}} ( q' + x_{Bj} P)^\mu$,  
and $J_\mu$ is the electromagnetic current.

\begin{figure}[h]
\centering{\ \psfig{figure=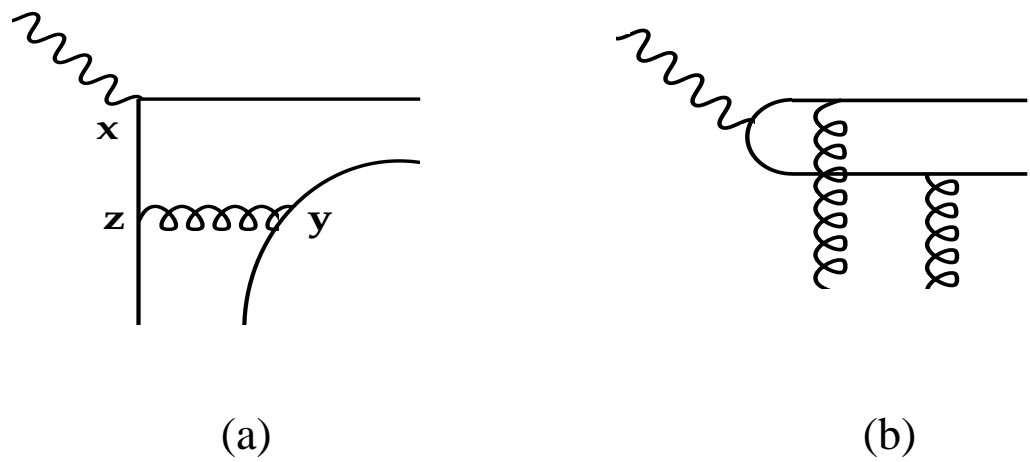,height=8cm,angle=-90}}
\caption{Typical graphs which contribute to the hard exclusive  
meson production amplitude.}
\label{fig:graphs}
\end{figure}

In second order perturbation theory we arrive at:
\begin{equation}
  S =  2 \, \pi \, i \, \alpha_s \, \int \! 
             d^4 x \, d^4 y \, d^4 z \, e^{-iq\cdot x} \,
             \langle M(q') N(P') | {\cal M}(x,y,z) | N(P) \rangle,
  \label{2order}  
\end{equation}
with the time-ordered product of quark and gluon 
fields:
\begin{equation}
{\cal M}(x,y,z) = T\left(\bar{\psi}^a(z) \gamma^{\rho} 
                   \psi^b(z) \bar{\psi}^c(x) 
                  \hat{\epsilon_L} \psi^c(x)
                  \bar{\psi}^d(y) \gamma^{\sigma} \psi^e(y)
                   A^A_{\rho}(z) \, A^B_{\sigma}(y)\right) 
                  t^A_{ab} t^B_{de}, 
\end{equation}
where $t^A$ and $t^B$ denote the generators of color SU(3).  As a next step we
perform the usual Operator Product Expansion of ${\cal M}(x,y,z)$.  Choosing,
for example, the flow of the hard momentum $q$ as in  
Fig.\ref{fig:graphs}(a), 
we obtain a term which can be
interpreted as photon-meson transition in a background quark field provided by
the nucleon. In the remaining (not shown) leading order contributions the
photon couples to one of the other possible quark lines.  A different choice of
the hard momentum flow results in a photon-meson transition in the background
gluon field as shown in Fig.\ref{fig:graphs}(b).  
This gluon contribution has
been discussed several times in recent papers 
\cite{Rad97,hood97,Zhy97}. We therefore
outline in the following the calculation of the quark part.  
Here the diagram in Fig.\ref{fig:graphs}(a)  corresponds in leading 
order $\alpha_s$ to:
\begin{equation}
{\cal M}(x,y,z) \ni 
\left[\bar{\psi}^d(y) \,\gamma^{\sigma} 
\,\psi^e(y) \right] 
\left[\bar{\psi}^c(x)                   
\hat{\epsilon_L}\, S^{q(ca)}(x-z)
\,\gamma^{\rho} 
\,\psi^b(z)\right]
S^{g(BA)}_{\sigma\rho}(y-z)\,
                  t^A_{ab} t^B_{de}.
\label{diagfig1}
\end{equation}
$S^{q}$ and $S^{g}$ stand for the quark and gluon perturbative Feynman 
propagators, respectively.
Next we perform a Fierz transformation in color- and Dirac-space.
Of course in color-space only the singlet piece survives  
when matrix elements between colorless hadronic initial and final states 
are taken. 
In Dirac-space we have:
\begin{equation} 
\left[\bar{\psi}^d(y) \,\gamma^{\sigma} 
\,\psi^e(y) \right] 
\left[\bar{\psi}^c(x)                   
\hat{\epsilon_L}\, S^{q(ca)}(x-z)
\,\gamma^{\rho} 
\,\psi^b(z)\right]
  = \sum_{i,j} C_{ij}^{\sigma\rho} \,{\hat O_i}(y,z)\, {\hat O_j}(x,y),
\label{Fierz}
\end{equation}
with the operators $\hat O_{i}$ and ${\hat O_j}$ defined
as:
\begin{equation}
\hat O_{i,j}(y,z) = 
\left[ \bar{\psi}(y) \, \Gamma_{i,j} \,\psi(z) \right].
\end{equation}  
The coefficients $C_{ij}$ are given by:
\begin{eqnarray}
&& C_{ij}^{\sigma\rho}  = -  \frac{1}{16} 
\mbox{Tr}\left[ \gamma^{\sigma}
\Gamma_j \,\hat{\epsilon}_L \,S^q(x-z) \,\gamma^{\rho} 
\Gamma_i \right],  
\nonumber \\
&&\mbox{with} \quad 
\Gamma_{i,j} \in \{ 1, \gamma_5,\gamma^{\mu},\gamma_5 \gamma^{\mu} , 
\sigma^{\mu\nu} \}.
\label{Cij}
\end{eqnarray}
For color singlet final states higher order corrections generate path-ordered
exponentials which ensure gauge-invariance of the operators $\hat O_{i,j}$
\cite{Rad97}.  As a next step leading twist pieces have to be
extracted.  In the massless limit, i.e. $P^2 = P'^2= q^{'2} = 0$, 
this can be
achieved by using the Sudakov decomposition of $\gamma$ matrices:
\begin{equation} \label{eq:Sudakov}
   \gamma^{\mu} = \frac{\hat{q'}}{P \cdot q'} \, P^{\mu}
                + \frac{\hat{P}}{P \cdot q'} \, q'^{\mu}
                + \gamma_{\perp}^{\mu}. 
\end{equation}
Typically only longitudinal terms 
proportional to $P$ or $q^\prime$ 
contribute at the twist-2 level. 
Evaluating the corresponding trace and
sandwiching operators ${\hat O}_i$ and ${\hat O}_j$ between initial and final
states gives the desired amplitudes in terms of  meson
distribution amplitudes and nucleon nonforward parton distributions.  An
important outcome of this calculation is that only operators with $\Gamma_i =
\Gamma_j$ contribute.  Since the spin and parity of the produced meson
determines the Dirac matrix in the meson amplitude, it automatically
fixes the involved nonforward distribution.

\subsection{Vector mesons} 

To leading twist accuracy two amplitudes for 
longitudinally and transversely polarized vector mesons exist 
(see \cite{BallBraun} for a recent discussion).
In the longitudinal case one has:
\begin{equation} \label{eq:meson_distr_VL}
  \langle M (q') |\, \bar{\psi}(x) \hat P \psi(y) \, | 0 \rangle
  =  q'\cdot P \,f_{V_L} \int_0^1 d\tau \, 
    \Phi_{V_L}(\tau) \,
    e^{i q' \cdot \, (\tau x + \bar{\tau} y)}, 
 \end{equation} 
where $f_{V_L}$ and $\Phi_{V_L}$ denote the 
corresponding decay constant and distribution amplitude.
According to our previous discussion  
this implies that the quark contribution to 
the production of longitudinally polarized vector mesons 
is determined by the nucleon matrix element 
$\langle N(P')| \bar{\psi}(z)\hat q'\psi(y) | N(P) \rangle$. 
As a consequence the nonforward quark distributions 
$F$ and $K$, defined in 
Eqs.(\ref{eq:dd_quark},\ref{def:corr_F_zeta}), enter.
The corresponding amplitude reads:
\begin{eqnarray}
{\cal A}^{q}_{V_L} & = & \, \pi \, \alpha_s \, 
           \frac{C_F}{N_c} \frac{1}{Q} 
           \frac{\bar{N}(P',S') \hat{q}' N(P,S)}{P \cdot q'}
      f_{V_L} \int_0^1 d\tau \frac{\Phi_{V_L}(\tau)}{\tau \bar{\tau}}
           \nonumber \\
         &\times&  \int_0^1 dX \left[ F_{\zeta}(X,t) + \bar{F}_{\zeta}(X,t)
             \right] \left( \frac{1}{X - i \epsilon}
             + \frac{1}{X - \zeta + i \epsilon} \right)
         \nonumber \\
         & + &  \, \pi \, \alpha_s \, \frac{C_F}{N_c} \frac{1}{Q} 
           \, \frac{\bar{N}(P',S') (\hat{q}' \hat{r} - \hat{r} \hat{q}')
           N(P,S)}{2 \, M \, P \cdot q'} f_{V_L}
           \int_0^1 d\tau \frac{\Phi_{V_L}(\tau)}{\tau \bar{\tau}}
           \nonumber \\
         &\times&  \int_0^1 dX \left[ K_{\zeta}(X,t) + \bar{K}_{\zeta}(X,t)
             \right] \left( \frac{1}{X - i \epsilon}
             + \frac{1}{X - \zeta + i \epsilon} \right).
\label{eq:vecm_quark}
\end{eqnarray}

Since the virtual photon and the final state vector meson 
carry the same C-parity, vector meson production can take place also in 
the gluon background field. Its contribution is 
\cite{Rad97,hood97,Zhy97}:
\begin{eqnarray}
{\cal A}^{g}_{V_L} & = & \, 2 \pi \, \alpha_s \, 
           \frac{1}{N_c} \frac{1}{Q} 
           \frac{\bar{N}(P',S') \hat{q}' N(P,S)}{P \cdot q'}
            \, \left( 1 - \frac{\zeta}{2} \right) \,
       f_{V_L} \int_0^1 d\tau \frac{\Phi_{V_L}(\tau)}{\tau \bar{\tau}}
           \nonumber \\
         &\times&  \int_0^1 dX \, \frac{G_{\zeta}(X,t)} 
             {(X - i \epsilon)
              (X - \zeta + i \epsilon)} + K-\mbox{terms},
\label{eq:vecm_gluon}
\end{eqnarray}
with $G$ being the nonforward gluon distribution (\ref{eq:dd_gluon})
of the nucleon. 

Transversely polarized vector mesons are described  by the amplitude
\cite{BallBraun}: 
\begin{equation} \label{eq:meson_distr_VT}
  \langle M (q') |\, \bar{\psi}(x) \sigma_{\mu\nu} \psi(y) \, | 0 \rangle
  = -i (\epsilon_{\mu} q'_{\nu} - \epsilon_{\nu} q'_{\mu}) 
    f_{V_{T}} \int_0^1 d \tau \, 
    \Phi_{V_T}(\tau) \,
    e^{i q' \cdot \, (\tau x + \bar{\tau} y)}. 
\end{equation}
Their production 
from longitudinally polarized photons involves chiral-odd nucleon 
nonforward parton distributions. 
Thus, in general, only the nonforward quark transversity distribution 
contributes. 
An explicit calculation shows, however, 
that in leading order $\alpha_s$ 
the twist-2 contribution
vanishes and one has to go either to higher
orders, or to higher-twists to obtain a non-zero result.

\subsection{Pseudoscalar mesons}

In the case of pseudoscalar mesons one ends up with 
the amplitude:
\begin{equation} \label{eq:meson_distr}
  \langle M (q') |\, \bar{\psi}(x)\gamma_5 \hat P \psi(y) \, | 0 \rangle
  = i \,q'\cdot P \,f_{PS} \int_0^1 d\tau \, 
    \Phi_{PS}(\tau) \,
    e^{i q' \cdot \, (\tau x + \bar{\tau} y)}, 
\end{equation} 
with the meson decay constant $f_{PS}$ and the 
distribution amplitude $\Phi_{PS}$. 
According to the above discussion 
the production process is sensitive to the nonforward nucleon matrix element 
$\langle N(P')| \bar{\psi}(z) \gamma_5 \hat q'\psi(y) | N(P) \rangle $. 
Thus the nonforward  polarized quark distributions 
$\Delta F$ and $\Delta K$ from 
Eqs.(\ref{eq:dd_quarkpol},\ref{def:corr_F_zeta}) enter. 
Collecting all leading order contributions gives   
for the pseudoscalar meson production amplitude:
\begin{eqnarray}
&&{\cal A_{PS}}  =  i \, \pi \, \alpha_s \, 
           \frac{C_F}{N_c} \frac{1}{Q} 
           \frac{\bar{N}(P',S') \gamma_5 \hat{q}' N(P,S)}{P \cdot q'}
    f_{PS} \int_0^1 d\tau \frac{\Phi_{PS}(\tau)}{\tau \bar{\tau}} 
           \nonumber \\
&& \hspace{0.5cm} \times \int_0^1 dX \left[ \Delta F_{\zeta}(X,t) - \Delta 
             \bar{F}_{\zeta}(X,t)
             \right] \left( \frac{1}{X - i \epsilon}
             + \frac{1}{X - \zeta + i \epsilon} \right) 
             +  K-\mbox{terms}, 
\label{eq:psm_quark}
\end{eqnarray}
Due to  C-parity conservation pseudoscalar meson production 
receives contributions only from C-odd nonforward distribution functions. 
As an
immediate consequence  two gluon exchange is impossible. 
At least three gluons have to be exchanged, which 
in the language of  twist expansion corresponds to  higher-twist.

\section{Results} 
\label{results}

Using the derived production amplitudes together 
with the model distributions from Sec.3 allows to 
calculate the production of  neutral pseudoscalar and vector mesons. 
In both cases we use the asymptotic meson distribution amplitude
\cite{BroLep,Rad}: 
\begin{equation}
\Phi_{PS}(\tau) = \Phi_{V_L}(\tau) = 6 \tau (1-\tau). 
\end{equation}
In numerical calculations we have neglected the practically unconstrained
"$K$-terms". As their contribution enters proportional to $r$ it is bound to
be small at small momentum transfers.

\subsection{ $\pi^0$ and $\eta$ production}

\begin{figure}[h]
\centering{\ \psfig{figure=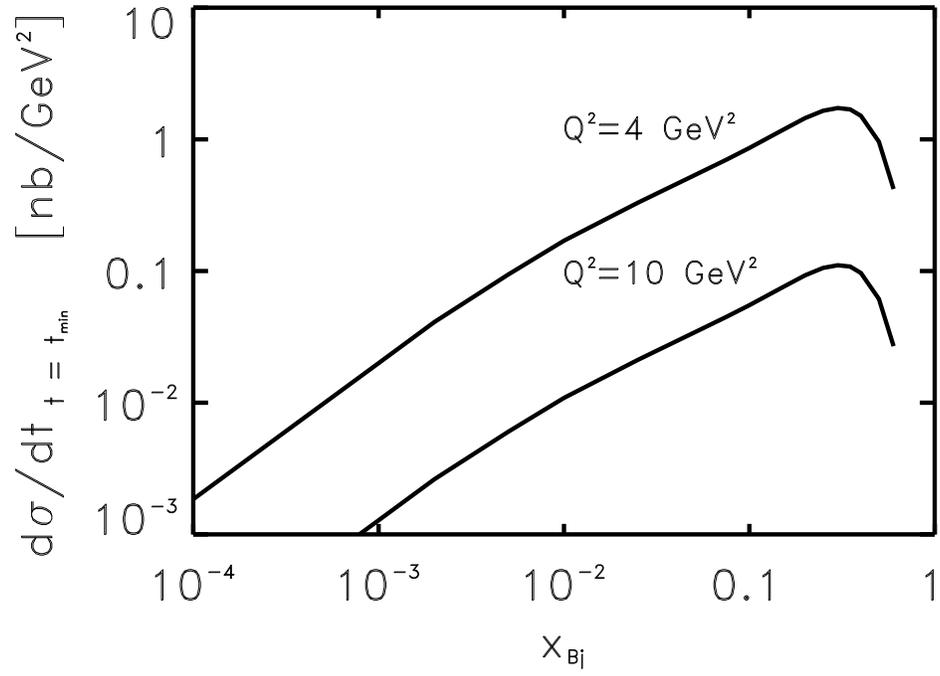,height=10.5cm}}
 \caption{Differential cross section for exclusive $\pi^0$ production 
from protons at $t = t_{min}$.}
\label{fig:pionfigure}
\end{figure}

To obtain predictions for a specific process the generic amplitudes 
from Sec.5 have to be furnished with flavor charges.
Using standard SU(3) wave functions for the pseudoscalar 
meson octet implies for  $\pi^0$ and $\eta$ production\footnote{We 
restrict our considerations to the pure octet state $\eta_8$, 
neglecting mixing with the singlet state $\eta_0$.}
an replacement of the nonforward distributions in 
Eq.(\ref{eq:psm_quark}) through distributions with specific flavor 
as listed in Tab.1.
\begin{table}[h]
\begin{center}
\begin{tabular}{|c|c|} \hline \rule [-3mm]{0mm}{8mm}
  & $\Delta F_{\zeta}(X,t) - \Delta \bar{F}_{\zeta}(X,t)$  \\
  \hline\hline
  $\rule [-3mm]{0mm}{8mm} \pi^0$ &  
  $ \sqrt{2} \left[  \frac{1}{3} \, 
   (\Delta u_{\zeta} - \Delta \bar{u}_{\zeta}) + 
   \frac{1}{6} \, (\Delta d_{\zeta} - \Delta \bar{d}_{\zeta}) 
   \right] $ \\ \hline
  $\rule [-3mm]{0mm}{8mm} \eta$ &  
  $\sqrt{6} \left[ \frac{1}{9} \, (\Delta u_{\zeta} - \Delta \bar{u}_{\zeta}) -
   \frac{1}{18} \, (\Delta d_{\zeta} - \Delta \bar{d}_{\zeta})
   + \frac{2}{9} \, (\Delta s_{\zeta} - \Delta \bar{s}_{\zeta})
   \right] $  \\ \hline
\end{tabular}
\label{tab:ps_flav}
\caption{Flavor structure for $\pi^0$ and $\eta$
  production. Analogous relations hold for the 
  $"K-\mbox{terms}"$.}
\end{center}
\end{table}
For $\pi^0$  production we use the standard value for the decay
constant  $f_{\pi} = 133$ MeV.  In Fig.\ref{fig:pionfigure} we present the
corresponding differential production cross section 
for a proton target taken at $t=t_{\min} = -x_{Bj}^2 M^2/(1-x_{Bj})$.  
We restrict ourselves to the region $|t_{min}| <
1$ GeV$^2$, which implies $x_{Bj} <  0.6$.

One finds that, up to logarithmic corrections, 
the production cross section drops as $1/Q^6$. 
Furthermore, at small values of $x_{Bj}$ the calculated cross 
section is proportional to  $x_{Bj}^{2 \lambda}$ with $\lambda \approx 0.6$. 
This behavior  is controlled by the small-$x_{Bj}$ behavior of the 
polarized valence quark distributions which enter in 
Eq.(\ref{def:nf_part_dist}). 
For the used parametrizations from Ref.\cite{GS96} 
one has indeed  
$x_{Bj}\Delta u_v(x_{Bj}) \sim  x_{Bj}\Delta d_v(x_{Bj}) 
\sim x_{Bj}^{\lambda}$. 
The decrease of the production cross section at large 
$x_{Bj}$ is due to the form factor in 
Eq.(\ref{def:nf_part_dist}). 
It is important to note 
that this behavior should be quite general and largely 
independent of  a  specific model for double distributions. 
The reason is that the rise of the production cross section 
with increasing $x_{Bj}$ is brought to an end through the 
decrease of the involved nonforward or double distributions 
at large momentum transfers $|t|$.
Since the latter is determined by a typical nucleon scale, 
$\Lambda \sim 1$ GeV, the maximum of the pseudoscalar meson 
production cross section at $t=t_{min}$ should occur when 
$t_{min}$ starts to become sizeable, say 
$-t_{min}/\Lambda^2 \gsim 0.2$. 
This implies $x_{Bj} \approx 0.3$ in accordance with 
our result. 

Our prediction for the $\eta$ production cross section  
turns out to be 
approximately a factor $2/3$ smaller than for $\pi^0$, 
but of similar shape. 
This is due to the comparable shape of the polarized 
$u$ and $d$ valence quark distributions from  Ref.\cite{GS96} 
which enter in Eq.(\ref{def:nf_part_dist}).

\subsection{Vector meson production}

For  $\rho^0,\Phi$ and $\omega$  production the nonforward 
distributions in Eq.(\ref{eq:vecm_quark},\ref{eq:vecm_gluon}) have to be 
modified according to Tab.2
\begin{table}[h]
\begin{center}
\begin{tabular}{|c|c|c|} \hline \rule [-3mm]{0mm}{8mm}
  & $F_{\zeta}(X,t) + \bar{F}_{\zeta}(X,t)$ & $G_{\zeta}(X,t)$ \\
  \hline\hline
  $\rule [-3mm]{0mm}{8mm} \rho^0$ &  
  $\sqrt{2} \left[ \frac{1}{3} \, (u_{\zeta} + \bar{u}_{\zeta}) + 
   \frac{1}{6} \, (d_{\zeta} + \bar{d}_{\zeta}) \right] $ 
   & $\frac{1}{\sqrt{2}} \,G_{\zeta}$  \\ \hline
  $\rule [-3mm]{0mm}{8mm} \omega$ &  
  $\sqrt{2} \left[ \frac{1}{3} \, (u_{\zeta} + \bar{u}_{\zeta}) -
   \frac{1}{6} \, (d_{\zeta} + \bar{d}_{\zeta}) \right] $ 
   & $\frac{1}{3 \sqrt{2}} \,G_{\zeta}$  \\ \hline
  $\rule [-3mm]{0mm}{8mm} \Phi$ &  
  $- \frac{1}{3} \, (s_{\zeta} + \bar{s}_{\zeta}) $& 
   $ - \frac{1}{3}\, G_{\zeta}$  \\ \hline 
\end{tabular}
\label{tab:vec_flav}
\caption{Flavor structure for $\rho^0$, 
         $\omega$ and $\Phi$ meson production.}
\end{center}
\end{table}
In the kinematic domain of small $x_{Bj} < 0.01 $ the quark part of the 
production amplitude (\ref{eq:vecm_quark}) is negligible. 
Then the well known SU(3) relation  for the  cross section ratios 
$\sigma(\rho):\sigma(\omega): \sigma(\Phi) = 9 : 1 : 2$ follows 
immediately from Tab.2.

In Ref.\cite{Rys93,FS94,FS96} hard exclusive neutral 
vector meson production has 
been investigated at small values of $x_{Bj}$ 
within the framework of perturbative QCD. 
In this work the nonforward  gluon  distribution 
which, in general, enters the production amplitude (\ref{eq:vecm_gluon}) 
has been approximated by the ordinary gluon distribution. 
In the limit of small $x_{Bj}$ the obtained amplitude corresponds to 
Eq.(\ref{eq:vecm_gluon}) after the 
replacement $G_{\zeta}(\zeta,t) \rightarrow \zeta \, g(\zeta)$, 
where $g$ denotes the ordinary gluon distribution.
However, note that for a typical behavior of the gluon  
double distribution $G(x,y,t)\sim x^{-\lambda}$ with 
$\lambda > 0$ one obtains from Eq.(\ref{eq:nonf_to_forw})
for the ratio: 
\begin{equation}
\frac{G_{\zeta}(\zeta,t=0)}{\zeta g(\zeta)} 
\approx 1 + \lambda \,\frac{\int_0^1 dy \,y \,G(\zeta,y,t=0)}
{ \int_0^1 \,dy \,G(\zeta,y,t=0)} > 1.
\end{equation}
For our model distribution from Sec.3 this  leads 
to an increase of the vector meson production cross sections  
by nearly a  factor of two as compared to the results of 
\cite{FS96}. 

It should be mentioned that 
the leading order calculations from Refs.\cite{Rys93,FS94,FS96}
overestimate the experimental data from HERA \cite{Crit97} 
by a typical factor $5-10$. 
Additional  Fermi motion corrections have been added to 
fit  experimental data. 
This emphasizes the need for a systematic investigation of 
higher twist effects \cite{Zhy97}.

\begin{figure}[h]
\centering{\ \psfig{figure=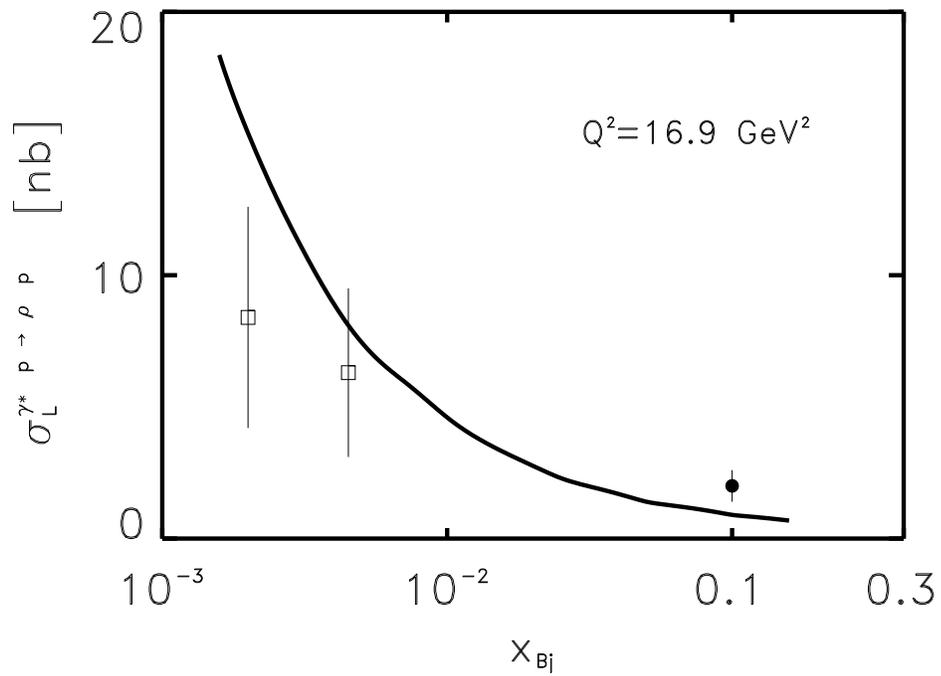,height=10.5cm}}
 \caption{Cross section for $\rho^0$ 
         production from a proton through the interaction of  
         longitudinally polarized photons. The data are taken  
         from ZEUS \cite{ZEUS} (open squares) 
         and NMC \cite{NMC94} (filled circles).}
\label{fig:rhofigure}
\end{figure}

In Fig.\ref{fig:rhofigure} we present the result for $\rho^0$ production 
from a proton in the 
kinematic domain of HERA \cite{ZEUS}  as obtained from the amplitudes in 
Eqs.(\ref{eq:vecm_quark},\ref{eq:vecm_gluon}), combined 
with the double distributions from Sec.3. 
For the  $\rho$ meson decay constant $f_{\rho} = 216$ MeV has 
been  used. 
After multiplying with a suppression factor 
$T(Q^2= 16.9\,\mbox{GeV}^2) \approx 0.1$ 
from \cite{FS96} qualitative agreement can be achieved.

Finally we would like to note that special attention should be paid 
to exclusive charged meson production. 
Here at large $Q^2$ nonforward parton distributions are probed 
which are related to matrix elements of twist-2 non-local operators 
off diagonal in flavor:
\begin{equation}
{\hat O}_{qq'}(x,y) = \bar q(x)\, \Gamma \,q'(y),
\end{equation}
with flavor dependent quark fields, $q=u,d,s$.  As a consequence hard
exclusive charged meson production offers possibilities to explore exotic
parton distributions which have not been measured elsewhere.  These obey
isospin symmetry relations analogous to the Bjorken sum rule known from 
polarized deep-inelastic scattering:
\begin{eqnarray}
\langle n | {\hat O}_{du}(x,y) | p \rangle & = &
\langle p | {\hat O}_{uu}(x,y) | p \rangle - 
\langle p | {\hat O}_{dd}(x,y) | p \rangle, \nonumber \\
\langle n | {\hat O}_{du}(x,y) | p \rangle & = &
\langle n | {\hat O}_{dd}(x,y) | n \rangle - 
\langle n | {\hat O}_{uu}(x,y) | n \rangle.
\label{IsoSR}
\end{eqnarray}
Here $\langle n |$ and $\langle p |$ denote neutron and proton
states, respectively. These sum rules imply a close relation of charged meson 
production cross sections. They allow, for example, to relate  
the isovector part of the amplitude for $\rho^0$ production to a linear
combination of $\rho^+$ and $\rho^-$ production amplitudes.

\section{Summary}

Ordinary parton distributions accessible e.g., in 
deep-inelastic scattering measure the  nucleon response 
to a process where one parton is removed and subsequently 
inserted back into the target along a light-like distance, 
without changing its longitudinal momentum. 
Generalized parton distributions, so-called nonforward 
distributions, can be studied in 
deeply virtual Compton scattering and hard exclusive leptoproduction 
of mesons. 
They describe a situation where the removed parton changes its
longitudinal momentum before returning to the nucleon. 
Furthermore,  hard exclusive charged meson production provides  
possibilities  to investigate  new processes where the 
removed quark carries different flavor than the returning one.

In this paper we have studied general properties of 
double  and nonforward parton distributions. 
We have found a symmetry of quark and gluon double distribution 
functions based on their relation to
nonforward matrix elements of QCD string operators. 
Its implication for phenomenological models has been  
outlined.

Although the leading order  QCD evolution equations seem to be more 
complicated in the non-forward than in the forward case, 
one can solve them explicitly using an expansion in a set of 
orthogonal polynomials instead of performing a Mellin integral in the  
complex plane of distribution function moments. 
We have outlined this method for the flavor nonsinglet case. 
A generalization  to  flavor singlet distributions is possible.

In the second half of this paper we have derived 
the amplitudes for  the hard exclusive production of neutral mesons
from longitudinally polarized photons. 
After suggesting a phenomenological model
for  double or nonforward distribution functions 
which obey  appropriate symmetries,  
we have presented results for exclusive 
$\pi^0$, $\eta$ and $\rho^0$ production.  

In the pseudoscalar case nonforward polarized 
valence quark distributions enter. 
Several features of the presented results   
should be independent of our  specific model for  nonforward distribution 
functions: 
at small $x_{Bj}$ pseudoscalar meson production cross 
sections drop  for decreasing $x_{Bj}$. 
This is related to a similar behavior 
of the corresponding forward distributions.
Furthermore, the production cross sections 
peak at moderate $x_{Bj}$. 
This is  due to the   
dependence of the involved nonforward distributions 
on the momentum transfer which should be  controlled 
by a typical nucleon scale. 

We also have presented results for $\rho^0$ production 
in the kinematic domain of recent HERA measurements. 
Here the production process is dominated by contributions 
from the nonforward gluon distribution.
For a large class of model distributions this tends to be 
larger than the corresponding forward distribution.

\bigskip
\bigskip

{\bf Acknowledgments}: 
We gratefully acknowledge discussions with W. Melnitchouk and A. Radyushkin.
This work was supported in part by BMBF,
KBN grant 2~P03B~065~10, and German-Polish exchange program X081.91.

\newpage


\newpage



\newpage


\newpage

\newpage


\end{document}